\documentclass[journal]{IEEEtran}

\usepackage{tikz}

\usepackage{amsmath}

\usepackage{epstopdf}
\usepackage{color, soul}
\usepackage{multirow}
\usepackage{tabularx}
\usepackage{soul}

\newcommand{\ave}[1]{\left\langle #1 \right\rangle}

\begin{document}
\newcolumntype{C}[1]{>{\centering\arraybackslash}p{#1}}
\title{Efficient Determination of Reverberation Chamber Time Constant}

\author{Xiaotian~Zhang,~Martin~P.~Robinson,~Ian~D.~Flintoft,~\IEEEmembership{Senior~Member,~IEEE},~John~F.~Dawson,~\IEEEmembership{Member,~IEEE}}
\markboth{Journal of \LaTeX\ Class Files,~Vol.~14, No.~8, August~2015}%
{Shell \MakeLowercase{\textit{et al.}}: Bare Demo of IEEEtran.cls for IEEE Journals}



\maketitle

\begin{abstract}
Determination of the rate of energy loss in a reverberation chamber is fundamental to many different measurements such as absorption cross-section, antenna efficiency, radiated power, and shielding effectiveness. Determination of the energy decay time-constant in the time domain by linear fitting the power delay profile, rather than using the frequency domain quality-factor, has the advantage of being independent of the radiation efficiency of antennas used in the measurement. 
However, determination of chamber time constant by linear regression suffers from several practical problems, including a requirement for long measurement times. Here we present a new nonlinear curve fitting technique that can extract the time-constant with typically 60\% fewer samples of the chamber transfer function for the same measurement uncertainty, which enables faster measurement of chamber time constant by sampling fewer chamber transfer function, and allows for more robust automated data post-processing. 
%Such properties of nonlinear curve fitting enables faster measurement of chamber time constant by sampling fewer chamber transfer function with the same frequency step, while not sacrificing as much accuracy as linear curve fitting does,
 Nonlinear curve fitting could have economic benefits for test-houses, and also enables accurate broadband measurements on humans in about ten minutes for microwave exposure and medical applications. The accuracy of the nonlinear method is demonstrated by measuring the absorption cross-section of several test objects of known properties. The measurement uncertainty of the method is verified using Monte-Carlo methods. 
\end{abstract}


% \fi
\IEEEpeerreviewmaketitle

\section{Introduction}

\IEEEPARstart{T}{he} properties of the reverberation chamber (RC) are described in detail by Hill \cite{hill2009electromagnetic}. 
As well as EMC and shielding effectiveness measurements \cite{IEC2011,holloway2007measuring}, RCs are widely used for the measurement of absorption cross-section (ACS), for characterisation of radio absorptive materials \cite{Hallbjorner2005Extracting}, and for biological studies \cite{melia2013broadband}. They are also used for communication channel simulation \cite{Dortmans2016}. In all of these applications knowledge of the chamber Q-factor or time constant is essential, and as the Q-factor depends on the chamber contents, it must be determined explicitly for each particular measurement undertaken. 

%In the frequency domain the Q-factor of an RC can be determined by measuring the coupling between two antennas \cite{hill2009electromagnetic}:
%\begin{equation}
%\label{Eq_QfactorFromG}
%Q = \frac{16 \pi^2 V}{\lambda^3} \frac{\ave{P_\mathrm{R}}}{\ave{P_\mathrm{T}}} 
%\end{equation}
%where $V$ is the chamber volume, $\lambda$ is the wavelength, $\ave{P_\mathrm{R}}$ is the average power received into a perfect antenna over a number of stirrer positions, and $\ave{P_\mathrm{T}}$ is the averaged power transmitted into the chamber. In a practical measurement the transmission, $S_{21}$, between the terminals of the two antennas is measured and the efficiency of the antennas used must be taken into account so that:
%\begin{equation}
%\label{Eq_GfromSparam}
%\frac{\ave{P_R}}{\ave{P_T}} =G=\frac{ \ave{\magn{S_{21}}^2}}{\eta_\mathrm{T}^\mathrm{int} \eta_\mathrm{R}^\mathrm{int}\left(1-\magn{S_{11}}^2 \right) \left( 1-\magn{S_{22}}^2 \right)}
%\end{equation}
%where $G$ is the mismatch-corrected average transfer function; $\eta_\mathrm{T}^\mathrm{int}$ and $\eta_\mathrm{R}^\mathrm{int}$ are the the radiation efficiencies of transmitting and receiving antennas; the terms $\left(1-\magn{S_{11}}^2 \right)$ and $\left( 1-\magn{S_{22}}^2 \right)$ are the mismatch efficiencies of the transmitting and receiving antennas, $S_{11}$ and $S_{22}$ being the measured, free-space, reflection coefficients of the antennas. 

The Q-factor, $Q$, and chamber time constant, $\tau$, at angular frequency $\omega$ are simply related by \cite{hill2009electromagnetic}:
\begin{equation}
\label{Eq_QfactorTimeConstant}
Q = \omega \tau \: .
\end{equation}

A common method for determining the chamber time constant is to do linear curve fitting on the power delay profile (PDP) on a logarithmic scale. The slope of the fitted straight line gives the rate of power loss in the RC, therefore the chamber time constant can be extracted from the slope. 
The biggest advantage of determining chamber time constant in this way is that the $\tau$ value is not sensitive to antenna radiation efficiency 
\cite{holloway2012reverberation}. 
The PDP is obtained by calculating the inverse fast Fourier transform (IFFT) of the scattering parameter $S_{21}$ measured in the frequency domain at the ports of two antennas in the RC \cite{ghassemzadeh2004measurement}. 
Since the time constant varies with frequency, a window function is used to select each particular frequency band
%\JFDins{, for the computation of the chamber time constant,}
 from a broadband $S_{21}$ measurement prior to the calculation of a PDP.

However, there are three difficulties in applying such a method. First, the windowed $S_{21}$ should have wide enough bandwidth containing enough frequency samples to give a PDP with high enough resolution for linear curve fitting, but dense sweeping $S_{21}$ is very time consuming, especially in wideband applications. 
Second, since the PDP is obtained from the IFFT of a windowed $S_{21}$ spectrum, the impulse response of the window function is convolved with PDP which distorts its shape%\cite{Archibald2002A, X2015On}
. 
Third, linear curve fitting does not give the correct chamber time constant in low signal-noise ratio (SNR) cases as explained in Section \ref{Sec_ACS_Det}.

In this paper, a new nonlinear curve fitting technique is presented for extracting chamber time constants more accurately and more efficiently. Compared to linear curve fitting, nonlinear curve fitting has two advantages. First, nonlinear curve fitting can cancel the window function's effect on the PDP, therefore the extracted chamber time constants are not affected by the specific choice of window function; second, since nonlinear curve fitting allows a narrower window function to be applied in the extraction of the chamber time constant, fewer samples of $S_{21}$ are required to be measured and the measurement time can be greatly reduced by a segmented sweep which samples $S_{21}$ only around desired frequency points. 

The accuracy of nonlinear curve fitting in determining chamber time constant is demonstrated by measuring the ACS of a several objects of known properties in the RC. The measurement speed is improved by continuous mode stirring and segmented frequency sweeping, which enables the ACS measurement at 171 frequencies to be completed in 11 minutes. The quick measurement speed also facilitates the study of measurement uncertainty. The type A uncertainty was obtained by repeating the ACS measurement 16 times, and the results were compared to the uncertainty predicted by the Monte-Carlo method. Good correspondence was observed between the measured uncertainty and the Monte-Carlo prediction. 

The remainder of this paper is divided into four sections.
In Section \ref{Subsec_Det_ACS} we review the the method of determining the chamber time constant and ACS in an RC. Section \ref{Sec_ACS_Det} shows the problems of extracting chamber time constant by linear regression and how the problems were solved by applying nonlinear fitting techniques.
Section \ref{Sec_Experiment} presents the validation experiments for the new measurement techniques.
%

% \section{Theory}
% \label{Sec_Det_ACS} 

\section{ACS measurement in an RC}
\label{Subsec_Det_ACS} 
The total average absorption cross section, $\ave{\sigma_\mathrm{tot}}$, of all lossy objects (including apertures) in an RC is defined as \cite{Gifuni2009On}:
\begin{equation}
\label{eqnACS}
\ave{\sigma_\mathrm{tot}}=\frac{\ave{P_T}}{S_c}\: ,
\end{equation}
where  $S_c$ is the power density in the chamber and $\ave{P_T}$ is the average power loss by all the objects in the RC. 
The average power $\ave{P_R}$ received by an antenna in the chamber has the following relation to $S_c$ \cite{hill2009electromagnetic}:
\begin{equation}
\label{eqnRX}
S_c=\frac{8 \pi \ave{P_R} }{\lambda^2}\: ,
\end{equation}
where $\ave{P_R}$ is the received power measured at the port of a lossless receiving antenna. 
Consider the Q-factor's relation to $\ave{P_R}$ and $\ave{P_T}$ in an RC \cite{hill2009electromagnetic}:
\begin{equation}
\label{Eq_QfactorFromG}
Q = \frac{16 \pi^2 V}{\lambda^3} \frac{\ave{P_\mathrm{R}}}{\ave{P_\mathrm{T}}} \: ,
\end{equation}
where $V$ is the volume of the RC. 
Equation \eqref{eqnACS} can be written as \cite{gradoni2012absorbing}:
\begin{equation}
\label{eqnACSQ}
\ave{\sigma_\mathrm{tot}}=\frac{\lambda^2 }{8 \pi}\frac{16 \pi^2 V}{Q\lambda^3}=\frac{2 \pi V}{Q\lambda}\: .
\end{equation}
Substituting \eqref{Eq_QfactorTimeConstant} into \eqref{eqnACSQ} gives:
\begin{equation}
\ave{\sigma_\mathrm{tot}}= \frac{V}{c\tau}\: ,
\label{eqnACStotaltau}
\end{equation}
where $c$ is the speed of light. 

The ACS of an object in an RC can % therefore % redundant
 be determined from the difference in $\ave{\sigma_\mathrm{tot}}$ for the chamber with and without the object.
From \eqref{eqnACStotaltau}  the ACS of a lossy object can be written as \cite{hill2009electromagnetic,tian2016measurement}:
\begin{equation}
\ave{\sigma_\mathrm{obj}}=\frac{V}{c}\left(\frac{1}{\tau_\mathrm{wo}}-\frac{1}{\tau_\mathrm{no}}
\right)\: ,
\label{eqnACSobjtau}
\end{equation}
where the subscript '$\mathrm{wo}$' means 'with object' loaded in the chamber; '$\mathrm{no}$' means 'no object' is loaded in the chamber. Equation \eqref{eqnACSobjtau} indicates an accurate ACS measurement relies on the accurate determination of chamber time constants. Therefore the validation of nonlinear curve fitting in extracting chamber time constant is demonstrated by measuring the ACS as in Section \ref{Sec_Experiment}.

\section{Determination of the chamber time constant}
\label{Sec_ACS_Det}
\subsection{Determining Chamber Time Constant by Linear Curve Fitting}
\label{Subsec_ACS_Time_Response}
\begin{figure}
\includegraphics[width=\linewidth]{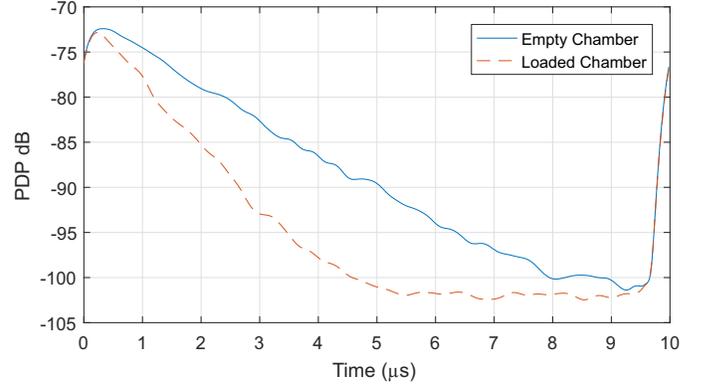}
\caption{An example of PDP measured in the University of York RC. The effect of loaded object can be seen from change of the slope of the  PDP.}
\label{Theory_Fig_PDPExampleLoadedUnloaded}
\end{figure}

The chamber time constant can be extracted from the power delay profile (PDP) where:
\begin{equation}
\label{Eqn_review_TimeResponse}
\text{PDP} = <|\text{IFFT}(S_{21}\cdotp W)|^2> \: .
\end{equation}
$W$ is a window function which is used to select the narrow frequency range required from a broadband measurement. We typically choose a set of window functions to calculate the time constant at each desired frequency from broadband measurement data. The PDP gives the power level in an RC as a function of time and typical results are shown in Fig. \ref{Theory_Fig_PDPExampleLoadedUnloaded},
which shows the PDP in decibels and how it changes as the chamber is loaded with a lossy object. The value of the chamber time constant can be obtained from the slope of the linear part of the PDP by curve fitting:
\begin{multline}
\text{PDP}_\text{dB}(t) = 10\text{log}_{10}(Ae^{-t/\tau})\\
= \left(-\frac{10\log_{10}e}{\tau}\right)t+10\text{log}_{10}A\: ,
\label{Theory_Eq_PDPLinear}
\end{multline}
where $\tau$ is the chamber time constant and $A$ is a positive constant which gives the signal power. Both $\tau$ and $A$ can be determined from linear curve fitting to the PDP on a dB scale. 
We call \eqref{Theory_Eq_PDPLinear} the linear model of PDP.

\begin{figure}
\includegraphics[width=\linewidth]{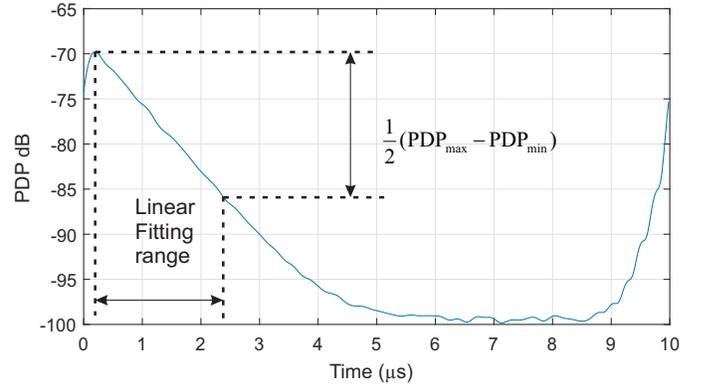}
\caption{The linear fitting range is chosen as the time interval which gives the upper half of PDP response. $\mathrm{PDP}_\mathrm{max}$ is the maximum reading of PDP; $\mathrm{PDP}_\mathrm{min}$ is the minimum reading of PDP.}
\label{Fig_Theory_LinearFittingRange}
\end{figure}

However, there are three problems in extracting the chamber constant by linear regression.
First, a suitable fitting range must be selected. 
The shape of PDP is not a perfect straight line but a combination of a declining slope and the  horizontal noise floor of the measurement system. 
In the method of \cite{cox1975distributions}, the linear fitting range was selected as the time interval that gives the top 30~dB of the PDP.
%, which is:
%\begin{equation}
%\{t_\mathrm{fit}\} \subseteq \left\{t \Big| \max\left[\mathrm{PDP}_\mathrm{dB}(t)\right]
% - \mathrm{PDP}_\mathrm{dB}(t) < 30 \mathrm{dB}\right\}  
%\end{equation}
%where $\{t_\mathrm{fit}\}$ denotes fitting range. 
However this fitting range only works well with a large signal noise ratio (SNR). In this study, the fitting range was chosen as the time range that corresponds to the top half of the PDP on a dB scale, as shown in in Fig. \ref{Fig_Theory_LinearFittingRange}.

\begin{figure}
\includegraphics[width=\linewidth]{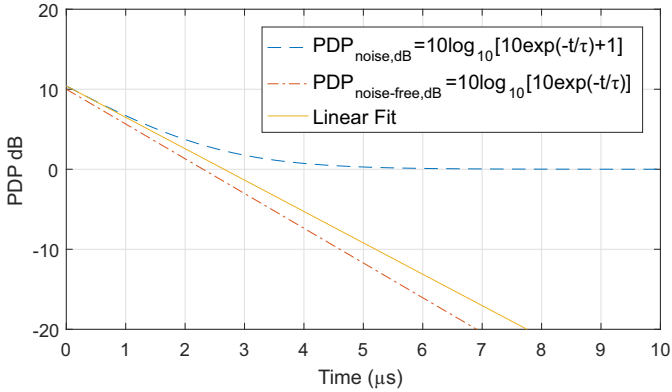}
\caption{Failure of linear curve fitting in low SNR case. The linear curve fit (solid line) does not match the original noise free PDP (dash-dot line). 
%No proper fitting range can be found on $t\in[{0 \mathrm{\mu s}, \, 10 \mathrm{\mu s}}]$ because the derivative of $\mathrm{PDP}_\mathrm{noise,dB}$ is always smaller than the slope of $\mathrm{PDP}_\mathrm{noise-free,dB}$. 
Fitting range was chosen by the method illustrated in Fig. \ref{Fig_Theory_LinearFittingRange}.}
\label{Fig_Theory_LowSNR}
\end{figure}

Second, in the low SNR case, the slope of the PDP is not a good indicator of the chamber time constant. 
This problem can be demonstrated by transforming the PDP model into a logarithmic scale and then calculating its derivative. 
The PDP with a noise floor can be modelled by \cite{zhang2016inverse}:
%Secondly, the linear curve fitting does not give an accurate evaluation on chamber time constant in low SNR cases. This can be demonstrated by calculating the derivative of the PDP in logarithmic scale. The PDP with noise floor can be modelled by \cite{zhang2016inverse}:
\begin{equation}
\mathrm{PDP}_\mathrm{noise,dB} = 10\log_{10}(Ae^{-t/\tau} + B)\: ,
\label{Eq_Theory_PDPwithNoise}
\end{equation} 
where $B$ is a positive constant that gives the noise power.
Calculating the derivative of $\mathrm{PDP}_\mathrm{noise,dB}$ with respect to $t$ gives:
\begin{equation}
\frac{\mathrm{d}\mathrm{PDP}_\mathrm{noise,dB}}{\mathrm{d}t}
=
\left(-\frac{10\log_{10}e}{\tau}\right)
\frac{(A/B)e^{-t/\tau}}{(A/B)e^{-t/\tau}+1}\: ,
\label{Eq_Theory_DiffPDPNoisedB}
\end{equation}
where $(A/B)$ is the SNR. 
%Since the fitting range is usually very close to $t=0$ as shown in Fig. \ref{Fig_Theory_LinearFittingRange}, substituting $t \approx 0$ into \eqref{Eq_Theory_DiffPDPNoisedB} gives:
%\begin{equation}
%\frac{\mathrm{d}\mathrm{PDP}_\mathrm{noise,dB}}{\mathrm{d}t} \Big| _{t\approx 0}
%\approx \frac{\mathrm{SNR}}{\mathrm{SNR}+1}\left(-\frac{10\log_{10}e}{\tau}\right)
%\label{Eq_Theory_DiffPDPApprox}
%\end{equation}
%If SNR is very high, $\mathrm{SNR}/(\mathrm{SNR}+1)$ is almost equal to 1, thus the derivative equals  
%the term in brackets in \eqref{Eq_Theory_DiffPDPApprox}. 
The term in the bracket is equal to the slope of noise-free PDP from which the correct chamber time constant can be obtained, as given in \eqref{Theory_Eq_PDPLinear}. 
If $(A/B)$ was a small value, the derivative in \eqref{Eq_Theory_DiffPDPNoisedB} would be dominated by the factors outside of the bracket, so  linear curve fitting would not give the right answer. 
As an example, the low SNR problem is illustrated in Fig. \ref{Fig_Theory_LowSNR} by setting $A=10,\, B=1, \, \tau=1\, \mu s$. 
The result of linear curve fitting does not correspond with the noise free PDP whose $A=10,\, B=0, \, \tau=1\, \mu s$. 
A SNR of $(A/B) \leq 10$ would make the problem even worse.

%In order to make the derivative of $\mathrm{PDP}_\mathrm{noise,dB}$ as close to the term in the bracket as possible, we should chose a fitting range in $t\geq0$ so that $(A/B)e^{-t/\tau}\gg1$. But such a linear fitting range is not findable when the value of $(A/B)$ is small because $(A/B)$ sets the upper bound of $(A/B)e^{-t/\tau}$ on $t\geq0$. The low SNR problem is demonstrated numerically in Fig. \ref{Fig_Theory_LowSNR} by choosing $A=10, B=1, \tau=1 \, \mathrm{\mu s}$ which gives $\mathrm{SNR=A/B=10}$. SNR lower than 10 will make the problem even worse.

%which is the slope of noise-free PDP. However, if SNR is very small, the derivative would be dominated by the value of $\mathrm{SNR}/(\mathrm{SNR}+1)$. This simple derivation shows that the slope of PDP in low SNR cases is not a good indicator of chamber time constant, therefore the linear regression is not reliable.

\begin{figure}
\includegraphics[width=\linewidth]{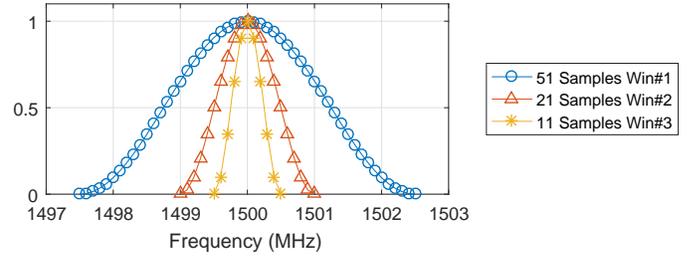}
\caption{Window functions at 15 GHz are all  raised cosine windows with rolling off factor $\beta=1$ and frequency step 100 kHz \cite{glover2010digital}.  }
%Win \#1: total width 5 MHz; Win \#2: total width 2 MHz; Win \#3: total width 1 MHz.}
\label{Fig_review_EffectofWindowFunctiononPDP}
\end{figure}

\begin{figure}[h]
\includegraphics[width=\linewidth]{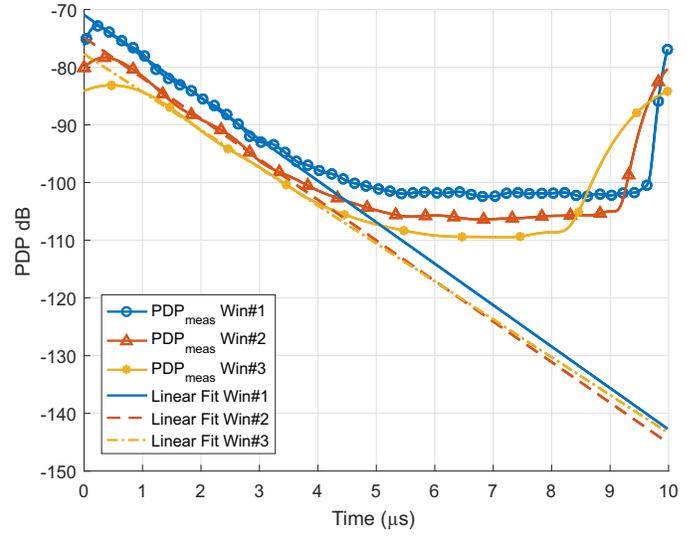}
\caption{PDP extracted by applying different window functions at 15 GHz. The filtered $S_{21}$ were all zero padded to zero frequency to show the effect of the window functions in full detail. The window function changes the shape of the PDP, thus the time constant given by linear regression is changed by the window function as well. The fitting range is selected in the way demonstrated in Fig. \ref{Fig_Theory_LinearFittingRange}.}
\label{Fig_review_EffectofFittingRangeonChamberTimeConstant}
\end{figure}

%Third, the wider main lobes of the impulse responses of narrower window functions would convolve with the PDP, thus the shape of PDP would be changed, which would eventually change the linear regression result, 
%%\JFDcom{due to the periodic nature of the IFFT, there is a rise in value at the end of PDP's time response,???} 
%as can be seen in Fig. \ref{Fig_review_EffectofWindowFunctiononPDP} and Fig \ref{Fig_review_EffectofFittingRangeonChamberTimeConstant}. 
%
Third, the multiplication of $S_{21}$ by a window function affects the shape of the PDP. The linear fit is quite sensitive to this, whereas the non-linear method includes the effect of the window function on its optimisation and so it is largely insensitive to the window used. In this paper we chose raised cosine windows, as shown in Fig. \ref{Fig_review_EffectofWindowFunctiononPDP}, because they give better results, compared to windows with a sharper roll-off, when the linear fit is used, though we have not exhaustively searched for an optimum window shape for the linear fit.
%
%The graph of Fig. \ref{Fig_review_EffectofWindowFunctiononPDP} shows three window functions of different widths at 15 GHz. The window functions are all full-roll-off raised cosine windows to minimize the  artefacts in the time domain. 
%We observe that the linear fit accuracy is quite sensitive to the window used and chose one that appears to give the best results though we have not search exhaustively. The nonlinear method is quite insensitive to the window used.
Fig. \ref{Fig_review_EffectofFittingRangeonChamberTimeConstant} 
compares the PDP calculated using the three different window functions and the same $S_{21}$ data set. %measured in the University of York RC, which is a galvanised steel chamber with dimensions of $4.7 \, \text{m} \times 3 \, \text{m} \times 2.37 \, \text{m}$. 
%15 GHz was chosen simply for easy demonstration of the problem, because the SNR is higher in the RC over 10 GHz and the noise floor can be observed on the graph of PDP. 
%Fig. \ref{Fig_review_EffectofFittingRangeonChamberTimeConstant} shows that the linear fit gives different results depending on the window chosen.    
It can be seen that the change in shape of the PDP due to the width of the window function has a significant effect on the linear fit.
%The linear fitting range was selected using techniques demonstrated in Fig. \ref{Fig_Theory_LinearFittingRange}.

\subsection{Determining the Chamber Time Constant by Nonlinear Curve Fitting}
The difficulty in extracting the chamber time constant from the PDP can be solved by introducing a nonlinear PDP model \cite{zhang2016inverse}. The new nonlinear model takes the effects of both the window function and the noise floor into account, therefore the chamber time constant can be extracted with better accuracy. 
This model is based on the assumption that the channel impulse response (CIR) can be modelled as the summation of many incoming rays with random phase shifts and exponentially decaying magnitudes \cite{saleh1987statistical}:
\begin{equation}
h(t) =
\sum_{l=0}^{\infty}
\beta_{l}
e^{j \theta_{l}}
\delta(t - T_l)\: ,
\label{Eq_Theory_MultiPathPulseResponse_FirstCluster}
\end{equation}
where $h(t)$ is the CIR; the coefficient $\beta_{l}$ is the magnitude of each ray, which decays exponentially with time; $e^{j\theta_l}$ is the phase shift of each ray; $\delta(t - T_l)$ is the Dirac delta function.
In terms of the central limit theorem, $h(t)$ observed at any specific moment in the chamber should follow a complex Gaussian distribution and its amplitude should decay exponentially as well:
\begin{equation}
\label{Theory_Eq_TimeDomainSingalNoNoise}
h(t) = V_s e^{-t / 2\tau} N_1(t) \: , 
\end{equation}
where $V_s$ is the received signal voltage; $N_1(t)$ is a standard complex Gaussian random process with zero mean and variance of one. 

Equation \eqref{Theory_Eq_TimeDomainSingalNoNoise} corresponds with Hill's idea that the transient behavior of an RC can be described by an exponential function \cite{hill2009electromagnetic}:

\begin{equation}
\label{Theory_Eq_TimeDomainExponentialModel}
U = U_s e^{-t/\tau}, \quad t> 0 \: , 
\end{equation}
where $U_s$ is a constant indicating the power density in the chamber.
However, \eqref{Theory_Eq_TimeDomainSingalNoNoise} still misses the effects of the noise floor and the window function. Adding both into  \eqref{Theory_Eq_TimeDomainSingalNoNoise} gives the filtered CIR:
\begin{equation}
\label{Theory_Eq_TimeDomainSignalWithNoise}
h(t) \otimes W(t) \ =  \left[ V_s e^{(-t / 2\tau)} N_1(t) + V_n N_2(t) \right] \otimes W(t) \: , 
\end{equation}
where $V_n$ is the background noise level; $N_2(t)$ is another standard complex Gaussian random process independent from $N_1(t)$; $W(t)$ is the time domain response of the window function; and $\otimes$ means circular convolution whose period equals the maximum time range of $h(t)$. 
%\eqref{Theory_Eq_TimeDomainSignalWithNoise} gives the complete form of nonlinear PDP model. 
%The effect of window function should also be added into the nonlinear PDP model. According to the convolution theorem, the response of filtered $h(t)$ can be calculated by convolving the original $h(t)$ with the time domain response of the window function. 
The power of the filtered $h(t)$ can be calculated, as in \cite{X2015On} (A brief proof can be found in Appendix \ref{Appendix_ProofofNonlinearModel}):
\begin{multline}
\label{Theory_Eq_PDPFilteredByWindow}
E\left(\text{PDP}(t_i)\right) = 
E\left(\left|h(t_i) \otimes W(t_i) \right|^2\right) = \\
\left[V_s^2 e^{-t_i / \tau} + V_n^2\right]
\otimes
\Big|W(t_i)
\Big|^2  \: , 
\end{multline}
where $E(\cdot)$ means expectation; $t_i$ is the $i$th sample of time in the time domain. Equation \eqref{Theory_Eq_PDPFilteredByWindow} is the full form of the nonlinear model for curve fitting. It is  controlled by four parameters: $V_s$, $V_n$, $\tau$, and $W$ in which $W$ is known. 
The model \eqref{Theory_Eq_PDPFilteredByWindow} can be fitted to the measured PDP using a method such as the Levenberg-Marquardt algorithm \cite{marquardt1963algorithm}.

The starting value for nonlinear fitting can be estimated in the following way. The initial value of $\tau$ is first estimated as $\tau_0$ by linear regression. Then we can generate a reference PDP signal $e^{-t_i/\tau_0} \otimes |W(t_i)|$ by which the starting value of $V_s$ can be determined as $V_{s,0}$ due to the linearity of convolution:
%\begin{equation}
\begin{multline}
\frac{\mathrm{PDP_{meas}}(t_i)}
{e^{-t_i/\tau_0}\otimes
\big|W(t_i)
\big|^2}
\approx
\frac{\left[V_s^2 e^{-t_i / \tau} + V_n^2\right]
\otimes
|W(t_i)|^2}{e^{-t_i / \tau_0}\otimes
\big|W(t_i)
\big|^2}
\\
\approx V_s^2 = V_{s,0}^2 \: ,
%\quad \left(\tau_0 \approx \tau, \, V_se^{-t_i / \tau} \gg V_n\right) \approx V_s^2\: , 
\label{Eq_Theory_NsStartingValue}
\end{multline}
%\end{equation}
where $\mathrm{PDP_{meas}}$ is the measured PDP response.  
Here $\mathrm{PDP_{meas}(t_i)} \approx E(\mathrm{PDP}(t_i))$ is assumed if the measured PDP is of good quality. 
%This is the linear approximation to the time constant is close to the real time constant ($\tau_0 \approx \tau$) and the PDP starts at a level much greater than the noise floor ($V_s \gg V_n$). 
We suggest calculating the value of \eqref{Eq_Theory_NsStartingValue} at the time when $\mathrm{PDP_{meas}}(t_i)$ reaches its maximum, because at this time the noise term $V_n^2$ can be neglected by assuming $V_s \gg V_n$.
%\JFDdel{What inside the bracket is the condition under which%\eqref{Eq_Theory_NsStartingValue} stands.} 
After the estimation of $\tau_0$ and $V_{s,0}$, the initial value of the noise, $V_{n,0}$, can be estimated. Here we use the reference signal $I(t_i) \otimes W(t_i)$ where $I(t_i)$ is a constant function whose value is 1 so that:
\begin{multline}
\frac{\mathrm{PDP_{meas}}(t_i) - V_{s,0}^2 e^{(-t_i/\tau_0)} \otimes W(t_i)}{I(t_i) \otimes W(t_i)}
\approx
\frac{V_n^2 \otimes W(t_i)}{I(t_i) \otimes W(t_i)} \\
= V_{n}^2 = V_{n,0}^2 \: , 
\end{multline}
where $V_{n,0}$ is the starting value of $V_n$.

%where $V_{n,0}$ is the starting value of $V_{n}$.

%Then a signal with unit signal strength can be generated as reference signal from which the starting value of $V_s$ can be determined as $V_{s,0}$ in term of the linearity of circular convolution:
%\begin{equation}
%V_{s,0} \approx ere
%\end{equation}
%The starting value of $V_s$ can be evaluated using the following property

%the starting value of $V_n$ was the minimum value of measured PDP.

\begin{figure}
\includegraphics[width=\linewidth]{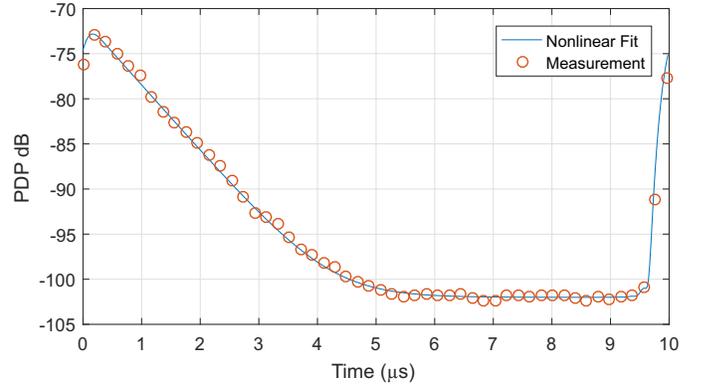}
\caption{A comparison between the nonlinear PDP model and PDP measured in the University of York reverberation chamber, with S21 filtered by Win \#1.}
\label{Theory_Fig_NonlinearModelPDP}
\end{figure}

\begin{figure}
\includegraphics[width=\linewidth]{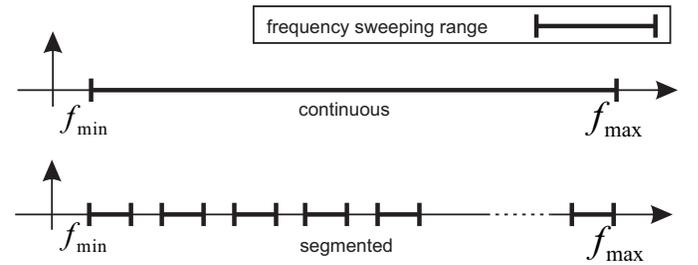}
\caption{Measurement speed can be increased by measuring only those S21 values included in the IFFT.}
\label{Theory_Fig_SegmentedSweeping}
\end{figure}

Fig. \ref{Theory_Fig_NonlinearModelPDP} shows that the optimized nonlinear model matches very well with the measured PDP. 
Compared to the linear regression for determining the chamber time constant, fitting with the nonlinear PDP model has the following advantages: 
First, the noise floor and window functions are parts of the nonlinear model, thus their effect can be compensated for in the determination of the chamber time constant. Second, since the effect of the window function is quantified in the nonlinear PDP model, a narrower window with fewer $S_{21}$ samples can be used in the determination of the chamber time constant. This may save measurement time because any $S_{21}$ values not included in the IFFT can be skipped in the measurement by segmented frequency sweeping, as illustrated in Fig. \ref{Theory_Fig_SegmentedSweeping}.

\subsection{Monte-Carlo Study on the Statistical Variance of Chamber Time Constant Determined by Nonlinear Curve Fitting}

\begin{figure}
\includegraphics[width=\linewidth]{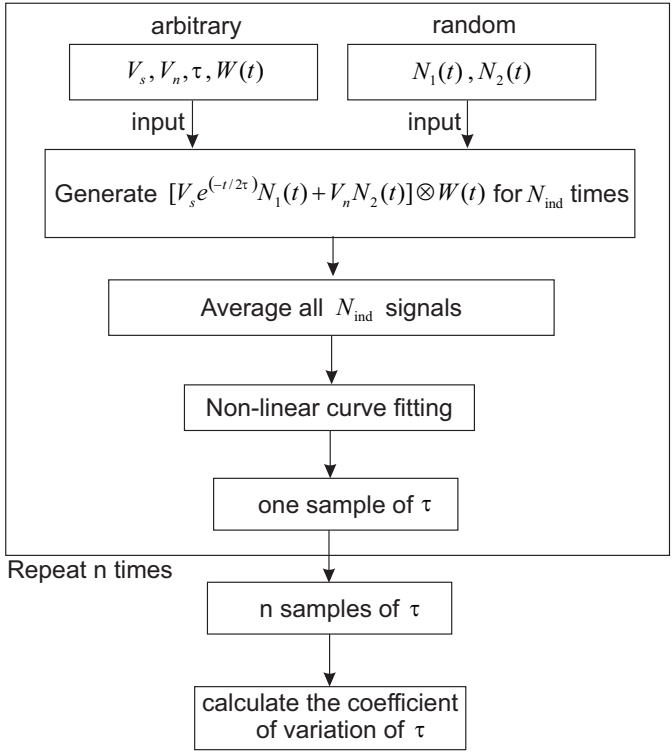}
\caption{Monte-Carlo method of estimating the measurement uncertainty.}
\label{Theory_fig_MonteCarloSimulation}
\end{figure}

Since the chamber time constant $\tau$ is extracted from the PDP whose statistical model is given in \eqref{Theory_Eq_TimeDomainSignalWithNoise}, the distribution of $\tau$ can be estimated by the Monte-Carlo method, as shown in Fig. \ref{Theory_fig_MonteCarloSimulation} \cite{bipm2009evaluation}.

The CIR model has the form of \eqref{Theory_Eq_TimeDomainSignalWithNoise}, therefore an artificial CIR can be generated with chosen values of $V_s$, $V_n$, $\tau$, and $W$; the sequence of Gaussian processes $N_1(t)$ and $N_2(t)$ were produced by the built-in function of MATLAB. Each generated CIR represents a single measurement of CIR at each independent stirrer position in the RC. Therefore the CIR measurement during stirrer movement can be simulated by generating the artificial CIR for $N_\mathrm{ind}$ times, where $N_\mathrm{ind}$ denotes the number of independent stirrer positions. Finally one $\tau$ value can be obtained by nonlinear curve fitting the PDP calculated from averaging the power of $N_\mathrm{ind}$ generated CIRs.

Such a process can be repeated for $n$ times to obtain $n$ values of $\tau$, then the distribution of $\tau$ can be calculated from these $n$ values.

\label{sec_montecarlo}

\section{Experiments}
\label{Sec_Experiment}
To demonstrate the accuracy of nonlinear curve fitting techniques for determining the chamber time constant, an ACS measurement on a lossy sphere was conducted in the University of York RC from 1 GHz to 16 GHz. 
The RC is a galvanised steel room with dimensions of $4.7 \, \text{m} \times 3 \, \text{m} \times 2.37 \, \text{m}$.
The transmitting and receiving antennas were respectively ETS 3115 and ETS 3117 horn antennas, which both work from 1 GHz to 18 GHz.  $S_{21}$ between two antenna ports
%\JFDcom{i and j being 2 and 1 ? or did you measure others for some reason ?}
 was measured by a vector network analyser (VNA). Segmented sweeping was applied to skip the frequencies not included in IFFT. The setup of frequency segments is as follows: The central frequencies of each segment are linearly stepped from 1~GHz to 16~GHz with a step size of 100~MHz, giving 151 segments in total; each segment is 5~MHz wide and each segment has 51 linearly distributed frequency samples. The segmented frequency sweeping from 1~GHz to 16~GHz was performed 800 times as the stirrer turned 360 degrees. The whole measurement took about 11 minutes.    
In general the frequency spacing of the points in each segment must be small enough to give a time response several ($\approx 5$) time constants long so that a good decay of the chamber energy occurs, and as the number of points in the segment determines the number of points in the time response, enough must be used to give a good representation of it.

\begin{figure}
\includegraphics[width=\linewidth]{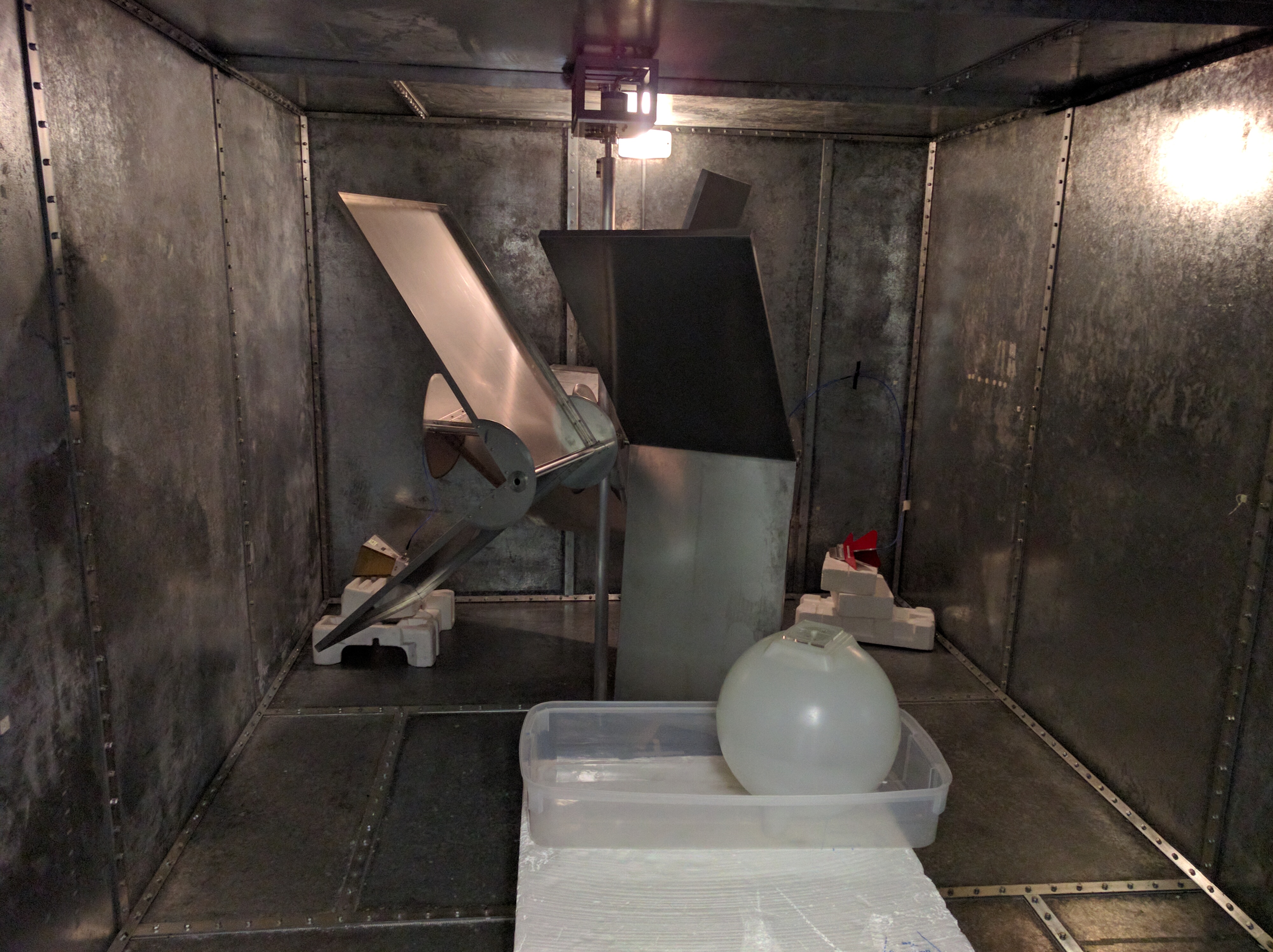}
\caption{Measurement of a sphere model in the University of York RC.}
\label{Experiment_Fig_ExperimentSetup}
\end{figure}

The sphere  under test is a spherical shell filled with deionized water (Fig. \ref{Experiment_Fig_ExperimentSetup}).  The outer radius of the sphere was 19.4~cm, obtained by measuring the circumference. The shell thickness was 3.9~mm, measured by a caliper close to rim of the sphere. The shell of the sphere is made of high density polyethylene (HDPE), whose relative permittivity is close to 2.35 over our frequency range \cite{riddle2003complex}. The complex permittivity of water was taken from Kaatze \cite{kaatze1989complex}. The room temperature was $20^\circ$C.

\begin{figure}
\includegraphics[width=\linewidth]{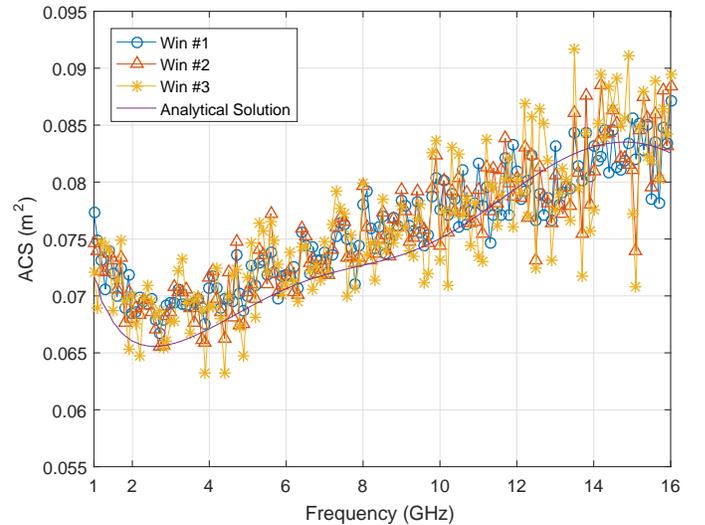}
\caption{The ACS of the sphere extracted by nonlinear fitting.}
\label{Experiment_Fig_ACSofSphereNonlinear}
\end{figure}

\begin{figure}
\includegraphics[width=\linewidth]{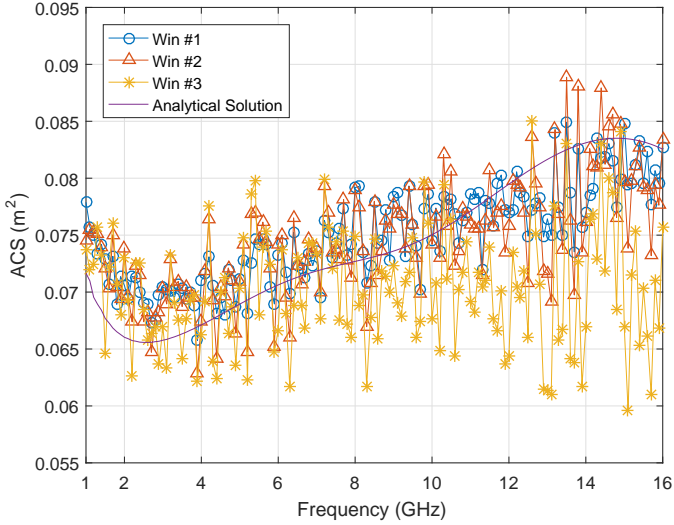}
\caption{The ACS of the sphere extracted by linear curve fitting.}
\label{Experiment_Fig_ACSofSphere}
\end{figure}

The three window functions shown in Fig. \ref{Fig_review_EffectofWindowFunctiononPDP} were applied to test the accuracy of  nonlinear curve fitting in extracting the chamber time constant. The ACS calculated from the chamber time constant is compared to the analytical solution given by Mie series calculator SPlaC V1.1 \cite{le2008splac} in Fig.  \ref{Experiment_Fig_ACSofSphereNonlinear}. The result for linear curve fitting is shown in Fig. \ref{Experiment_Fig_ACSofSphere}.

Fig. \ref{Experiment_Fig_ACSofSphere} shows that the linear curve fitting loses accuracy when narrower window functions are applied.  The worst case happens when the window function is only 1 MHz wide (Win \#3). In this case the measured ACS extracted by linear curve fitting is 10\% lower than the analytical solution. 
The mean absolute percentage errors (MAPEs) of the ACSs extracted by linear curve fitting with Win \#1, Win \#2, Win \#3 are 4.0\%, 5.0\% and 8.5\%. The MAPE is defined as \cite{tofallis2015better}:
\begin{equation}
\mathrm{MAPE}(\sigma_\mathrm{meas}) = \mathrm{mean}(
\Big|
\frac{\sigma_\mathrm{meas}(f) - \sigma_\mathrm{sim}(f)}{\sigma_\mathrm{sim}(f)}
\Big|) \times 100\%  \: , 
\end{equation} 
where $\sigma_\mathrm{meas}$ is measured ACS of the object under test; $\sigma_\mathrm{sim}$ is the theoretical value of ACS; and $\mathrm{mean}(\cdot)$
denotes averaging over frequencies from 1 GHz to 16 GHz. 

 The nonlinear curve fitting achieves a much better accuracy in determining the chamber time constant, thus a more accurate ACS was obtained, as shown in Fig. \ref{Experiment_Fig_ACSofSphereNonlinear}. The MAPEs of the ACSs extracted by nonlinear curve fitting with Win \#1, Win \#2, Win \#3 are 3.4\%, 3.5\% and 4.6\%. Compared to linear curve fitting with Win \#1, the nonlinear curve fitting with Win \#2 gives the ACS with better accuracy but from 30 fewer samples of $S_{21}$. Even in the case of applying a 1-MHz width window, the measured ACS result still follows the analytical solution, only with a larger variance.  

%We observe that the linear fit accuracy is quite sensitive to the window used and chose one that appears to give the best results though we have not search exhaustively. The nonlinear method is quite insensitive to the window used.

%Fig. \ref{Experiment_Fig_ACSofSphereNonlinear} and Fig. \ref{Experiment_Fig_ACSofSphere} show linear fitting is sensitive to the window function, and the shape/roll-off factor of seems to give the minimal error, though we have not exhaustively searched all shapes whilst the nonlinear fitting is not.

%\begin{figure}
%\includegraphics[scale=0.5]{Figures/FDandTDMehothod.eps}
%\caption{Comparison between IFFT and FD method}
%\end{figure}

\begin{figure}
\includegraphics[width=\linewidth]{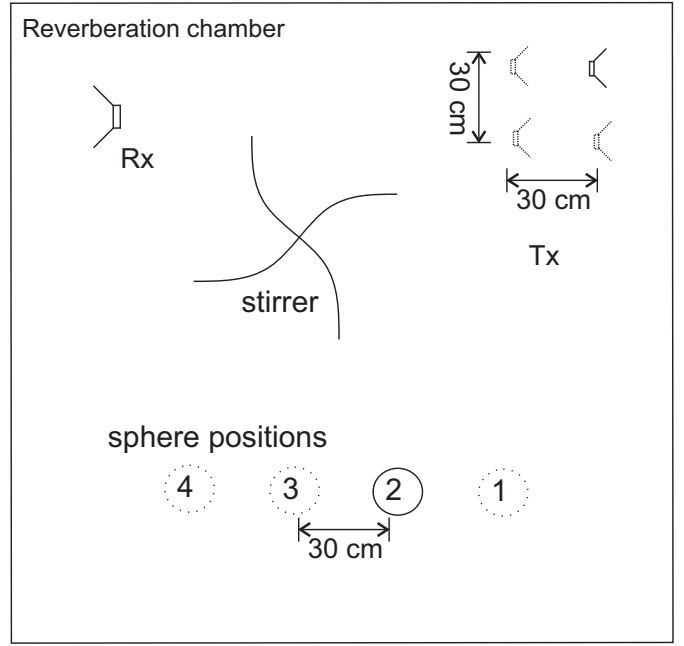}
\caption{The uncertainty study set up: the transmitting antenna was moved to 4 positions, and for each antenna position, the sphere was moved to 4 different positions too.}
\label{Fig_Experiment_SetupsUncertainty}
\end{figure}

\begin{figure}
\includegraphics[width=\linewidth]{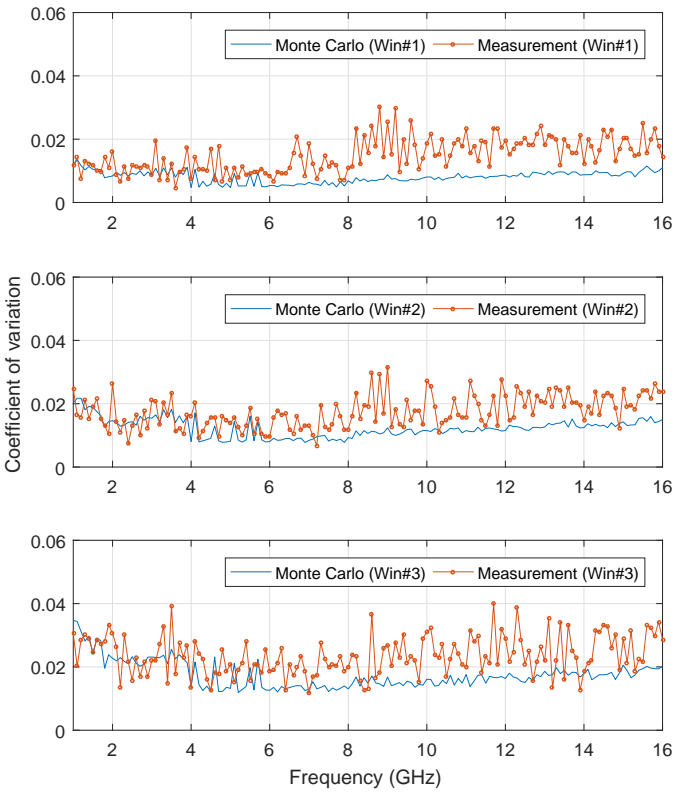}
\caption{The coefficient of variation of ACS given by measurement and by the Monte-Carlo method. Since the Monte-Carlo method only estimates statistical uncertainty of the multipath model, the discrepancy at high frequencies may be due to other sources such as imperfect stirring, moving of cables, etc.}
\label{Experiment_Fig_StandardDeviationofRelativeDifference}
\end{figure}

To evaluate the uncertainty of measured ACS extracted from nonlinear curve fitting, a series of measurements with similar setups, but different positioning of the transmitting antenna and the sphere model, were performed. A simple diagram of measurement setups is shown in Figure \ref{Fig_Experiment_SetupsUncertainty}. The receiving antenna was moved to 4 different positions, at least one wavelength apart (30 cm) from each other to ensure field independence. The sphere was also moved to 4 different positions for each receiving antenna position, which gives 16 different measurement setups in total. 
The nonlinear curve fitting with window function $\mathrm{Win} \#1$ was used to extract ACSs from the 16 measurements and the measurement uncertainty was characterized by calculating the coefficient of variation of 16 ACS results. The coefficient of variation is defined as the ratio of the standard deviation to the mean \cite{everitt2002cambridge}. The measurement uncertainty was also evaluated in the same way with the application of $\mathrm{Win} \#2$ and $\mathrm{Win} \#3$. The coefficient of variation obtained from measurement was compared to that given by the Monte-Carlo model in Fig. \ref{Experiment_Fig_StandardDeviationofRelativeDifference}. The figure shows that the Monte-Carlo model can successfully predict the measurement uncertainty and that the application of narrower windows tend to give higher uncertainty in evaluating ACS. 

\begin{figure}
\includegraphics[width=\linewidth]{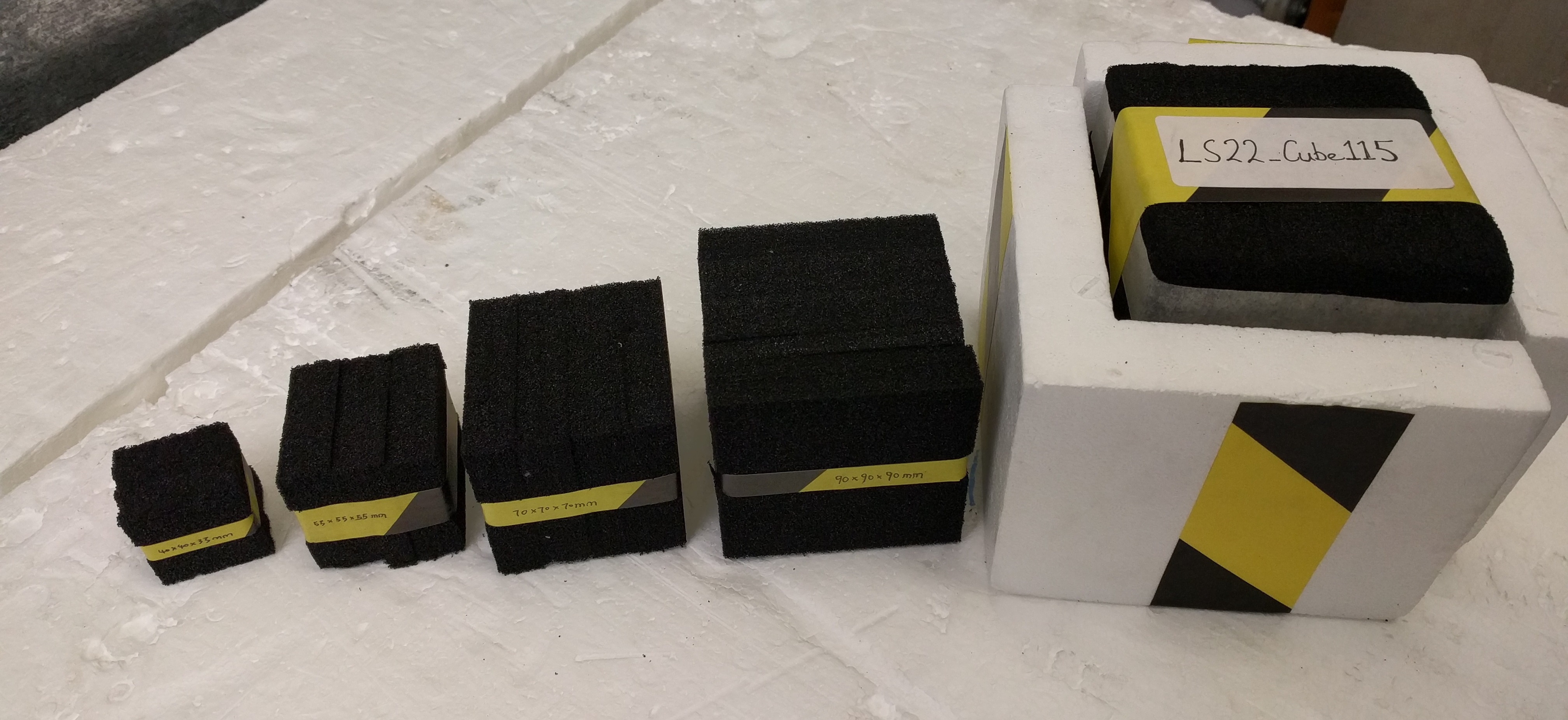}
\caption{The series of absorber cubes made from carbon-loaded foam.}
\label{Experiment_Fig_CubePic}
\end{figure}

To further test the measurement range of the nonlinear curve fitting technique, the ACS of a series of cubes (and one cuboid), fabricated from LS22 absorber \cite{LS22Data} of different sizes, were measured (Fig. \ref{Experiment_Fig_CubePic}). The complex permittivity of LS22 absorber was fitted to a three-pole Debye dispersion model \cite{flintoft2016measurable}:
\begin{equation}
\hat{\epsilon}_r = \epsilon_\infty 
+
\sum_{k=1}^3 \frac{\Delta \epsilon_k}{1+j\omega\tau_k} + \frac{\sigma_\mathrm{DC}}{j\omega\epsilon_0}  \: , 
\end{equation}
where $\epsilon_\infty=1.1725$, $\Delta\epsilon_1=1.04\times10^{-3}$, $\Delta\epsilon_2=17.9$, $\Delta\epsilon_3=0.490$, $\tau_1=55.3 \mathrm{ms}$, $\tau_2=0.188 \mathrm{ns}$, $\tau_3=6.20 \mathrm{ps}$, and $\sigma_\mathrm{DC}=0.1\mathrm{mS/m}$.
The ACS of the cubes in an RC was simulated by the CST time domain solver with the method of Carlberg \cite{carlberg2004calculated}. 

The MAPEs of the measured cube ACSs are listed in Table \ref{Tab_Cubemeas}.
\begin{table}
\centering
\caption{Mean absolute percentage error (MAPE) of cube ACS measurement.}
\label{Tab_Cubemeas}
\begin{tabular}{C{1.89cm}|C{0.4cm} C{0.9cm}|C{0.4cm} C{0.9cm}|C{0.4cm} C{0.9cm}}
\hline
\multirow{2}{*}{cube/cuboid size} & \multicolumn{2}{c|}{Win \#1} & \multicolumn{2}{c|}{Win \#2} & \multicolumn{2}{c}{Win \#3} \\
\cline{2-7}
 & linear & nonlinear & linear & nonlinear & linear & nonlinear \\
\hline
40*40*33 $\mathrm{mm}^3$   & 58\% & 27\%& 95\% & 36\%& 144\% & 50\%\\
 $(50\mathrm{mm})^3$  & 31\% & 19\%& 44\% & 23\%& 60\% & 28\% \\
 $(70\mathrm{mm})^3$   & 26\% & 18\%& 38\% & 18\%& 49\% & 21\%\\
$(90\mathrm{mm})^3$ & 16\% & 11\%& 21\% & 10\% & 28\% & 13\% \\
$(115\mathrm{mm})^3$   & 8\% & 5\%& 11\% & 6\%& 15\% & 7\%\\
\hline
\end{tabular}
\end{table}
The table shows that the cube ACS extracted by nonlinear curve fitting is more accurate than that obtained by linear curve fitting in all experiments, no matter which window function was used. 
Even though the accuracy of the ACS measurement deteriorated as the size of cube became smaller, the ACS of the smallest absorber was still able to be determined with a MAPE of 27\%, which was achieved by measuring only  51 $S_{21}$ samples about each desired frequency and applying nonlinear curve fitting.

%\begin{figure}
%\includegraphics[width=\linewidth]{Figures/CubeACS.eps}
%\caption{The ACS of absorber cubes measured in the reverberation chamber. Solid lines are measurement data, dashed lines are corresponding numerical simulation results}
%\label{Experiment_Fig_CubeACS}
%\end{figure}

\label{Experiment}

\section{Conclusion}
A new non-linear fitting method has been demonstrated, which allows accurate automated calculation of the chamber time constant from the PDP of band-limited IFFT data from a reverberant environment, without knowledge of the antenna efficiencies. It overcomes the problems of measurement noise floor and frequency window effects on the PDP data that make the linear fitting technique unreliable. This allows a fast segmented frequency sweep to be used to determine the chamber time constant over a wide frequency range. The operation of the method has been validated by comparison of the ACS of a spherical test object with that computed by means of the Mie series and with a range of absorptive cubes in comparison with a solution from a full wave solver. The ACS extracted by nonlinear curve fitting shows better accuracy in all of the validation experiments compared to that given by linear curve fitting. Combined with the use of mode stirring and a segmented frequency sweep, it significantly reduces the test time for measurements, which has been most useful, particularly with human subjects where a large group study is involved and a short test time is important. A Monte-Carlo model for the prediction of the accuracy of the non-linear fitting method has also been presented and validated against measurement.

% or


\appendices
\section{Proof for the nonlinear model}
Assume the signal received at the port of the receiving antenna in the time domain has the form of \eqref{Theory_Eq_TimeDomainSignalWithNoise} which is:
\begin{multline*}
h(t) = h_s(t) + h_n(t) = V_s e^{(-t / 2\tau)} N_1(t) + V_n N_2(t)  \: , 
\end{multline*}
where $h_s(t) = V_s e^{(-t / 2\tau)}N_1(t)$ and $h_n(t)=V_n N_2(t)$, the subscripts 's' and 'n' means 'signal' and 'noise'; $V_s$ and $V_n$ are the signal level and noise level, which are real numbers; $N_1(t)$ and $N_2(t)$ are two independent complex Gaussian random processes with zero mean and variance of one. 
Written in discrete form:
\begin{equation}
h(m) = h_s(m) + h_n(m)\\  \: , 
\label{Appendix_Eq_DiscreteFormofSignalandNoise}
\end{equation}
where
\begin{gather}
h_s(m) = V_s \exp{\left(\frac{-m \Delta t}{2\tau}\right)} N_1(m) \: , \\ 
h_n(m) = V_nN_2(m)  \: , 
\label{Appendix_Eq_DiscreteFormhsandhn}
\end{gather}
%h(m) = V_s e^{(-m \Delta t/ 2\tau)} N_1(m) + V_n N_2(m)
%where $h_s(m) = V_s e^{(-m \Delta t/ 2\tau)} N_1(m)$ and $h_n(m) = V_nN_2(m) $
$m$ is the index of responses in the time domain and $\Delta t$ is the time step size.

According to the properties of the discrete Fourier transform, the signal filtered (multiplied) by a window function in the frequency domain equals the circular convolution of their response in the time domain, therefore \eqref{Appendix_Eq_DiscreteFormofSignalandNoise} filtered by a window function can be written as:
\begin{multline}
h(m)\otimes W(m) = h_s(m)\otimes W(m)  \\
+ h_n(m) \otimes W(m)  \: , 
\label{Appendix_Eq_CircluatedConvolutionofhsandhnwithwindow}
\end{multline}
where $W(m)$ is the impulse response of the window function in the time domain. It is obtained from doing the IFFT on the spectrum of window function $W(f_k)$ zero-padded all the way to zero frequency.

In real measurement, the power response of \eqref{Appendix_Eq_CircluatedConvolutionofhsandhnwithwindow} is:

\begin{multline}
\left|h(m)\otimes W(m)\right|^2 =\\
\left|h_s(m)\otimes W(m)\right|^2 +
\left|h_n(m)\otimes W(m)\right|^2 + \\
\overline{\left[h_s(m)\otimes W(m)\right]}\left[h_n(m)\otimes W(m)\right] + \\
\left[h_s(m)\otimes W(m)\right]\overline{\left[h_n(m)\otimes W(m)\right]} \: , 
\label{Appendix_Eq_PowerResponseFiltered}
\end{multline}
where the bar over a term $\overline{\alpha}$ means complex conjugate of $\alpha$. Then the expectation of \eqref{Appendix_Eq_PowerResponseFiltered} is calculated. Because of the independence between $N_1(m)$ and $N_2(m)$, the two rightmost terms of \eqref{Appendix_Eq_PowerResponseFiltered} vanish:
\begin{multline}
E\left(\left|h(m)\otimes W(m)\right|^2\right) =
E\left(\left|h_s(m)\otimes W(m)\right|^2\right) +\\
E\left(\left|h_n(m)\otimes W(m)\right|^2\right) \: . 
\label{Appendix_Eq_ExpectationofPowerResponse}
\end{multline}
Due to the property of Gaussian random process that the random variables at any two different moments are independent, \eqref{Appendix_Eq_ExpectationofPowerResponse} can be simplified as:

\begin{equation}
E\left(\left|h(m)\otimes W(m)\right|^2\right)
=
\left[V_s^2 e^{-t / \tau} + V_n^2\right]
\otimes
|W(m)|^2 \: , 
\label{Appendix_Eq_NonlinearModel}
\end{equation}
which is  \eqref{Theory_Eq_PDPFilteredByWindow}.

\label{Appendix_ProofofNonlinearModel}

%\section{}

%% use section* for acknowledgment
%\section*{Acknowledgment}


\ifCLASSOPTIONcaptionsoff
  \newpage
\fi



%\bibliographystyle{IEEEtran}
\bibliographystyle{IEEEtran}
\bibliography{IEEEabrv,citations}

%\begin{thebibliography}{1}
%\end{thebibliography}

% needed around the contents of th
%e optional argument to biography to prevent

%\begin{IEEEbiography}{Xiaotian Zhang}
%\end{IEEEbiography}


%\newpage



%\vfill


\end{document}